\documentclass[aps,prl,twocolumn,showpacs,superscriptaddress,groupedaddress,floatfix]{revtex4}  
\usepackage{graphicx}  
\usepackage{dcolumn}   
\usepackage{bm}        
\usepackage{amssymb}   
\usepackage{color}

\def\apj{Astrophys. J.}
\def\apjl{Astrophys. J. Lett.}

\def\aap{Astron. Astrophys. }

\def\mnras{Mon. Not. Roy. Astron. Soc. }

\def\nat{Nature}
\def\prl{Phys. Rev. Lett.}
\def\prd{Phys. Rev. D.}

\def\lp{\left(}
\def\rp{\right)}
\def\be{\begin{equation}}
\def\ee{\end{equation}}
\def\ba{\begin{eqnarray}}
\def\ea{\end{eqnarray}}
\def\go{\mathrel{\raise.3ex\hbox{$>$}\mkern-14mu
             \lower0.6ex\hbox{$\sim$}}}
\def\lo{\mathrel{\raise.3ex\hbox{$<$}\mkern-14mu
             \lower0.6ex\hbox{$\sim$}}}

\begin{document}

\title{Resonant Shattering of Neutron Star Crusts}
\author{David Tsang}\email{dtsang@tapir.caltech.edu} \affiliation{TAPIR, California Institute of Technology, Pasadena, CA, USA}
\author{Jocelyn S. Read}\email{jsread@relativity.phy.olemiss.edu} \affiliation{Department of Physics and Astronomy, University of Mississippi, Oxford, MS, USA}
\author{Tanja Hinderer} \affiliation{TAPIR, California Institute of Technology, Pasadena, CA, USA}\affiliation{Department of Physics, University of Maryland, College Park, MD, USA}
\author{Anthony L. Piro} \affiliation{TAPIR, California Institute of Technology, Pasadena, CA, USA}
\author{Ruxandra Bondarescu}\affiliation{Department of Physics, Pennsylvania State University, State College, PA, USA}\affiliation{Institute for Theoretical Physics, University of Zurich, Switzerland}

\date{\today}

\begin{abstract}
The resonant excitation of neutron star (NS) modes by tides is investigated as a source of short gamma-ray burst (sGRB) precursors. We find that  the driving of a crust-core interface mode can lead to shattering of the NS crust,  liberating $\sim 10^{46}-10^{47}$ erg of energy seconds before the merger of a NS-NS or NS-black hole binary. Such properties are consistent with Swift/BAT detections of sGRB precursors, and we use the timing of the observed precursors to place weak constraints on the crust equation of state. We describe how a larger sample of precursor detections could be used alongside coincident gravitational wave detections of the inspiral by Advanced LIGO class detectors to probe the NS structure. These two types of observations nicely complement one another, since the former constrains the equation of state and structure near the crust-core boundary, while the latter is more sensitive to the core equation of state.
\end{abstract}
\pacs{97.60.Jd, 97.80.-d, 98.70.Rz, 95.85.Sz}
\maketitle

{\it Introduction.} Short-hard gamma-ray bursts (sGRBs) are intense bursts
of gamma-rays with isotropic equivalent energies of $E_{\rm iso} \sim
10^{50}-10^{51}$ erg, and durations $ < 2$\,s \citep{nakar:07a}. The
leading model for the progenitors of sGRBs is the merger of a neutron star
(NS) with another NS or the tidal disruption of a NS by a black hole (BH)
\citep{Paczynski1986,Goodman1986}. Studying the response of NSs 
in dynamic systems provides a window into the physics of their crust and
core, probing material up to many times nuclear density 
in states inaccessible to terrestrial laboratories.  

Flares occurring before the main GRB---precursor flares---were recently
identified for a handful of sGRBs \citep{Abdo2009, Troja2010}. In
particular, precursors $\sim1$--$10$\,s prior to the main flare were detected  with high significance for three sGRBs out of the 49 considered in \citet{Troja2010}. Until these discoveries, precursor flares had only been identified prior to some Long Gamma Ray Bursts (e.g. \cite{Burlon2009, Koshut1995}). Detailed sGRB precursor mechanisms remain largely unexplored, but since the precursors occur when the NSs are strongly interacting, they present an exciting opportunity to learn about the interior or region around the NSs. Magnetospheric interaction has been proposed as one explanation for the precursors, but is unlikely to be sufficiently strong unless the magnetic fields exceed magnetar strength \citep{Hansen2001}. Direct tidal crust cracking has also been suggested \citep{Kochanek1992b, Penner2011}. The crust cracks when the crust breaking strain $\epsilon_b \simeq 0.1$ \citep{Horowitz2009b} is exceeded by the direct tidal ellipsoidal deformation \citep{Damour2009a}
\begin{equation}
	\frac{\delta R}{R} \sim 0.1\frac{h_1}{0.8} \frac{q}{1+q} \frac{R_{12}^3}{M_{1.4}}\left(\frac{f_{gw}}{10^3 \,{\rm Hz}}\right)^2 \left(1+ PN\right),	\label{eq:deform}
\end{equation}
where $h_1$ is the quadrupole shape Love number of the deformed body,
$M_{1.4}=M/1.4\ M_{\odot}$ and  $R_{12}=R/12$\ km are its mass and radius,
respectively, $q$ is the mass ratio of the binary, and $f_{gw}$ is the
gravitational wave frequency (twice the orbital frequency); 
post-Newtonian corrections to the tidal potential ($PN$) are found in \cite{Vines2011}. At the frequency when $\delta R/R\sim \epsilon_b$, the gravitational wave inspiral timescale \citep{MTW}, is 
\be
	t_{gw} = \frac{f_{gw}}{\dot{f}_{gw}}
	= 4.7\times10^{-3}\,{\rm s}\left(\frac{\cal M}{1.2 M_\odot}\right)^{-5/3} \left(\frac{f_{gw}}{10^3\, {\rm Hz}} \right)^{-8/3},
	\nonumber
	\\
	\label{eq:trr}
\ee
where ${\cal M} = M_1^{3/5}M_2^{3/5}/(M_1 + M_2)^{1/5}$ is the chirp mass. Since this timescale is so short, unless the GRB emission is delayed for seconds after the objects coalesce (not generally expected), direct tidal deformation cannot explain the majority of observed precursors. 

In this letter we present an alternate tidal mechanism that can shatter the
crust and cause flares: the excitation of a resonant mode by periodic tidal
deformation. We identify a particular mode that has an
amplitude concentrated at the crust-core interface, and show that its
coupling to the tidal field is sufficient to fracture the crust. Further
driving of this mode can continue to cause fractures and deposit energy
into seismic oscillations until the elastic limit is reached, releasing the entire elastic energy of the crust as it shatters. Such an event may be observable as a precursor flare. 

\citet{Steiner2009} showed that the identification of
quasiperiodic oscillations in magnetar flares with toroidal shear modes can constrain physical parameters such as nuclear symmetry
energy and we similarly demonstrate how multi-messenger probes of sGRB precursor bursts
can be used to measure the interface mode frequency and constrain the crust equations
of state.   

{\it Analysis.}  The tidal excitation of NS normal modes in a binary has been well studied for fluid modes (e.g. \citep{Lai1994, Shibata1994}), and Reisenegger and Goldreich \citep{Reisenegger1994} hinted at possible associated electromagnetic signatures.  Including the solid crust of a NS into the mode analysis changes the normal mode spectrum significantly \citep{McDermott1988}. In addition to the fluid modes of the core (e.g. f-modes, g-modes), the solid crust induces several new modes, including crustal shear modes and interface modes between the crust and core.

For the present study we focus attention on the mode most likely to be tidally excited and subsequently crack the crust. The toroidal crust shear modes have large shear strains, but do not couple significantly to the spheroidal tidal field and are negligibly excited. Low order core g-modes can couple to the tidal field \citep{Lai1994}, but they do not penetrate the crust. The f-mode of the NS has the strongest tidal coupling, and can strongly perturb the crust. However, the f-mode frequencies are typically $\go10^3$\, Hz. These frequencies are not reached by the binary before merger, or are in resonance only at late times (see eq. [\ref{eq:trr}]). This would make any observable signal of this mode difficult to separate from the sGRB emission. In contrast to these other modes, the $l=2$ spheroidal crust-core interface mode (first identified by \citep{McDermott1988}), or i-mode, is ideal for resonant crust fracture as it couples significantly to the tidal field and has a large shear strain near the base of the crust. In addition, its frequency is low enough ($\sim 100$\,Hz) for tidal resonance to be well separated from the merger ($t_{gw}\simeq2\ {\rm s}$ at $f_{gw}\simeq100\ {\rm Hz}$ using eq. [\ref{eq:trr}]).

\begin{figure}
\includegraphics[width=\columnwidth]{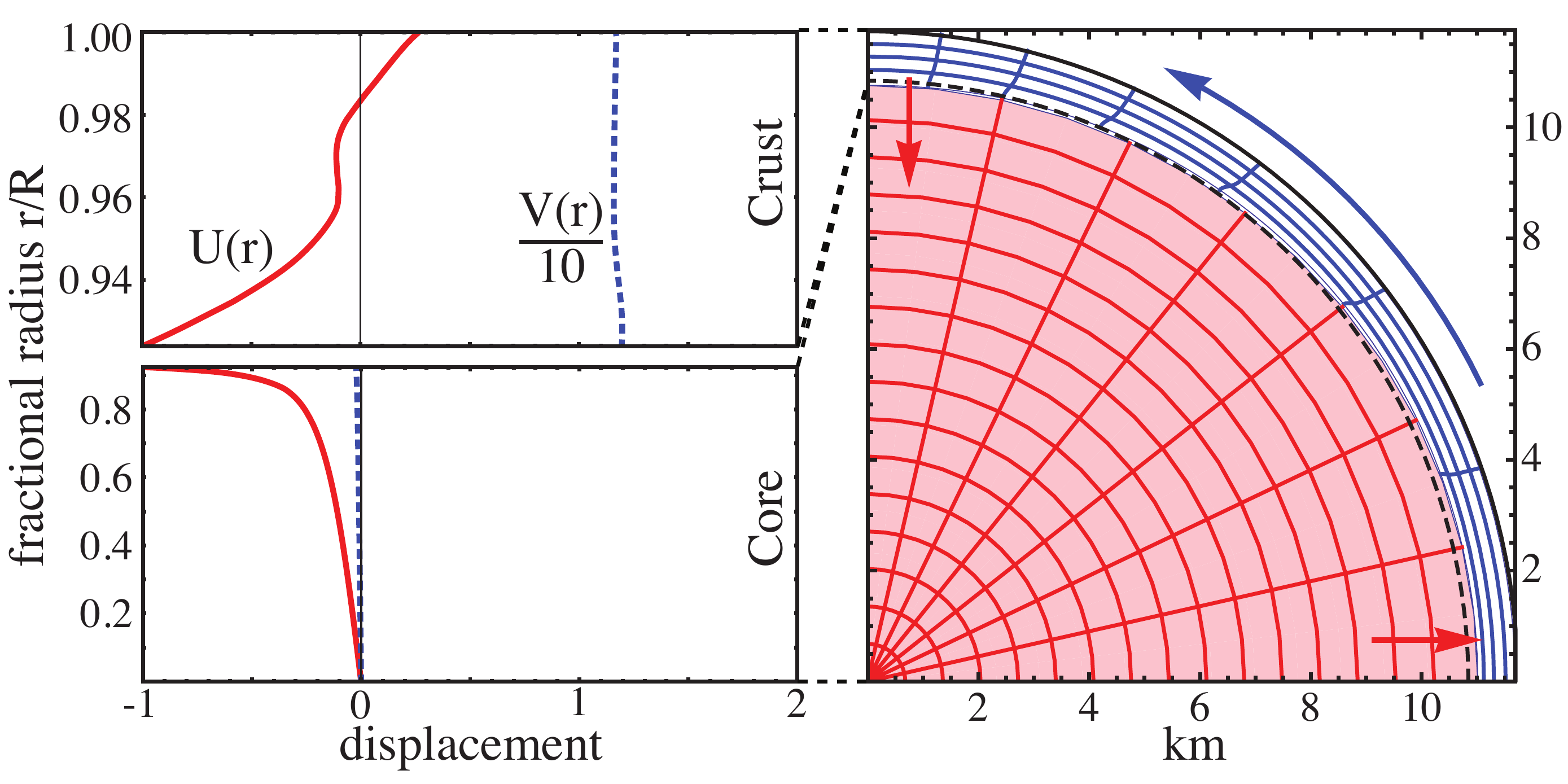}
\caption{\label{fig:eigenfunction} Left Panels: The $l=2$ crust-core i-mode eigenfunctions as a function of radius where the Lagrangian displacement is $\xi(r, \theta, \phi) = U(r) {\rm Y}_{2\pm2}(\theta, \phi) \boldsymbol{\hat{r}} + V(r) \boldsymbol{\nabla} {\rm Y}_{2\pm2}(\theta, \phi)$.  A $1.4 M_{\odot}$ $12$\,km NS with SLy4 \citep{Steiner2009} equation of state is used for the background model. Right Panel: Cross section of the NS deformation due the mode. The binary companion is located along the horizontal axis, about which the deformation has rotational symmetry. The fluid core is ellipsoidally deformed towards the binary companion. The solid crust is sheared strongly and compressed to make up for the deformation of the core such that the outer surface is only slightly perturbed radially. The horizontal strain and radial deformation peak at the base of the crust, making excitation of this mode ideal to crack the crust. The solid black line denotes the surface of the unperturbed star, while the dashed black line denotes the unperturbed crust/core transition radius.}
\end{figure}

We construct background models of spherical neutron stars with the Oppenheimer-Volkoff equations. For
the discussion below, we use the $1.4 M_{\odot}$ (12\,km) NS with Skyrme Lyon (SLy4) EOS
\citep{Steiner2009} and a core/crust transition at baryon density of
$n_b\sim 0.065$\,fm$^{-3}$ as a fiducial case. Taking the NS perturbation
equations from \citep{McDermott1988,Reisenegger1994} we solve for the
crust-core i-mode with a crust shear modulus $\mu$ given by
\citep{Strohmayer1991} (which ignores free neutron interactions,
and superfluid entrainment effects),
\be
\mu = \frac{0.1194}{1 + 0.595(173/\Gamma)^2}\frac{n_i(Ze)^2}{a}~,
\ee
where Z is the atomic number of the ions, $n_i$ is the ion density, $\Gamma \equiv (Ze)^2/ak_bT$ is the ratio of the Coulomb to thermal energies and $a = (3/4\pi n_i)^{1/3}$ is the average inter-ion spacing. 
We assume $ T \approx 10^8$\,K and ignore any magnetic field  since a field strength $\go10^{15}$G is
required for the magnetic energy density to be comparable to the shear
modulus $\mu\sim 10^{30}$\,erg\,cm$^{-3}$ at the base of the
crust, and
such a field has decay time that is likely much shorter than the binary
lifetime of the system \citep{Harding2006}. The exact mode frequency and
eigenfunction depend on properties near the crust-core boundary including
EOS, crust
shear modulus and crust-core transition density. There is also a weaker
dependence on bulk properties of the star such as its mass. For this model
the i-mode frequency is found to be $f_{\rm mode} \simeq 188$\,Hz, with
eigenfunction shown in Fig. \ref{fig:eigenfunction}. 

As the binary inspirals, it sweeps through successively higher frequencies until it passes through a resonance with the mode frequency. The timescale of during which the tidal driving is approximately phase coherent with the i-mode is estimated with a random phase approximation to find $t_{\rm res} \sim \sqrt{t_{gw}/\pi f_{gw}}$ \citep{Lai1994},
\be
t_{\rm res}  \sim 
 8\times10^{-2} \, {\rm s} \left(\frac{ {\cal M} }{1.2 \, M_\odot} \right)^{-5/6}
\left(\frac{ f_{\rm mode} }{100 \, {\rm Hz}} \right)^{-11/6}~,
\ee
where we have used the fact that $f_{\rm mode}\approx f_{gw}$ during the resonance. During this time, we estimate the degree of resonant excitation of the i-mode using the formalism of \citet{Lai1994}. From this we calculate the dimensionless overlap integral between the $l=2$ i-mode and the tidal field finding
 \be
Q \equiv\frac{1}{M R^2}\int d^3 x \rho~ \boldsymbol{\xi}^* \cdot 
\boldsymbol{\nabla}[ r^2 {\rm Y}_{2\pm2}(\theta, \phi)]\simeq 0.041~,
\label{eq:q}
\ee
where $\boldsymbol{\xi} \equiv U(r,t) {\rm Y}_{2\pm2} \boldsymbol{\hat{r}} + 
V(r,t) \boldsymbol{\nabla} {\rm Y}_{2\pm2}$ is Lagrangian displacement mode eigenvector normalized such that $\int \,d^3x\,\rho\, \boldsymbol{\xi} \cdot \boldsymbol{\xi}^*= MR^2$, and ${\rm Y}_{2\pm2}(\theta, \phi)$ is the spherical harmonic with $l=2$ and $m=\pm2$ with both $m$ modes excited equally, and the numerical factor on the right-hand side is from our calculation using the SLy4 equation of state.

During the resonance, the energy of the i-mode increases quickly over the timescale $t_{\rm res}$. The maximum possible  energy the mode can attain was explored by \citep{Lai1994}, ignoring crust fracture, orbital back reaction and other non-linear effects. Using equation (6.11) from this work, we estimate the maximum possible energy of the i-mode to be
\be
	E_{\rm max} \simeq 5\times10^{50}~ {\rm erg}\, f_{188}
	^{1/3} Q_{0.04}^2 M_{1.4}^{-2/3}R_{12}^2 ~q\left( \frac{2}{1+q}\right)^{5/3}
\ee  where $f_{188} \equiv f_{\rm mode}/188$\,Hz and $Q_{0.04} \equiv Q/0.04$. 
We next compare this to the mode energy needed to create a break in the crust $E_{\rm b}$. For a given amplitude mode, we can numerically calculate the dimensionless strain $\epsilon$ (see e.g. \citep{Landau}), and identify the location of maximum strain within the crust. In general, this is found at the base of the crust and concentrated near the NS equator. We numerically find the minimum amplitude $|\boldsymbol{\xi}_{\rm b}|$ needed for $\epsilon = \epsilon_{\rm b}$ at some location within the crust, which in turn implies a minimum energy,
\ba
	E_{\rm b} =  (2 \pi f_{\rm mode})^2 \int d^3x\, \rho \, \boldsymbol{\xi}^*_{\rm b} \cdot \boldsymbol{\xi}_{\rm b}
	 \simeq 5\times10^{46} {\rm erg} \,\epsilon_{0.1}^2
	 \label{eq:eb}
\ea
where $\epsilon_{0.1} \equiv \epsilon_{\rm b}/0.1$. Since $E_{\rm b}\ll
E_{\rm max}$ the i-mode will be amplified enough to break the crust
early in the resonant energy transfer.

These initial breaks have a characteristic size similar to the
thickness of the crust $\Delta r\sim 0.1 R$, and thus release elastic
energy $\sim \epsilon_b^2 \mu \Delta r^3\simeq 10^{43}\ {\rm erg}$ from the
excited mode. This energy is converted primarily into a broad
spectrum of seismic waves, peaked at characteristic frequency $\sim
(\mu/\rho)^{1/2}/(2\pi \Delta r) \sim 200$\,Hz. Seismic waves at these low
frequencies do not couple efficiently to the magnetic field, so the energy
remains mostly in the crust \citep{Blaes1989}. But the fracture is not
limited to a single event; the crust
continues to fracture as the location of maximum strain moves around the
star, since the NS is not tidally locked \citep{BildstenCutler1992} and the
crust heals extremely quickly \citep{Horowitz2009b}. 
Seismic energy injected by these fractures can build up in the crust until it reaches the elastic
limit and the crust shatters. The elastic limit  can be estimated as
\ba E_{\rm elastic} \equiv \oint \epsilon_b^2 \mu ~dx^3 \sim 4\pi R^2
\epsilon_b^2\mu \Delta r\sim2\times10^{46}\ {\rm erg}\,\epsilon_{0.1}^2.
\ea
%
The crust fractures cause the i-mode amplitude to saturate 
at amplitude $|\boldsymbol{\xi}_{\rm b}|$, with the tidal energy
transfer rate \citep{Lai1994}
\begin{eqnarray}
	\dot{E}_{\rm tidal} = \int d^3x \rho \frac{\partial \boldsymbol{\xi}^*_b}{\partial t} \cdot \boldsymbol{\nabla} \Phi_g,
\end{eqnarray}
where $\Phi_g = \sum_{lm} \Phi_{lm}$ is the gravitational potential due to the companion. We have the components $\boldsymbol{\nabla}\Phi_{2\pm2} = (f_{\rm gw}/\pi)^2q(q+1)^{-1}W_{2\pm2} \boldsymbol{\nabla}(r^2Y_{2\pm2})$, where $W_{2\pm2}=(6\pi/5)^{1/2}$ and taking $\partial \boldsymbol{\xi}^*_b/\partial t = 2\pi f_{\rm mode}\boldsymbol{\xi}^*_b$, we can use equations (\ref{eq:q}) and (\ref{eq:eb}) to rewrite this expression as
\ba
	\dot{E}_{\rm tidal} &\approx& \lp\frac{3\pi}{40} \rp^{1/2}(2\pi f_{\ gw})^2 M^{1/2}R Q E_{\rm b}^{1/2}\frac{q}{q+1},
	\nonumber
	\\
	&\approx& 10^{50}\ {\rm ergs\ s^{-1}} f_{188}^2E_{46}^{1/2}Q_{0.04},
\ea
where $E_{46}\equiv E_b/10^{46}\ {\rm erg}$. 
This high rate implies that only a small fraction of the resonant time, $E_{\rm elastic}/\dot{E}_{\rm tidal}\sim 10^{-3}\ {\rm s}\ll t_{\rm res}$,  is required for the energy in the crust to build up to the elastic limit. 

When this limit is reached and the crust shatters, this scatters the elastic energy and mode energy to higher frequency oscillations, which can couple efficiently to the magnetic field \citep{Thompson1995, Thompson1998}. If emission occurs over the observed precursor timescale this implies a luminosity $L \sim (E_b + E_{\rm elastic})/0.1\,{\rm s} \sim 10^{48}$ erg s$^{-1}$. If the magnetic field is strong enough that the perturbations are linear, $B \gg 10^{13}\,{\rm G}\, R_{12} (L/10^{48} {\rm erg\,s}^{-1})^{1/2}$, the energy propagates as Alfv{\'e}n waves along open magnetic field lines, and the resultant emission will be non-thermal. For weaker fields the magnetic perturbations are highly non-linear and can generate strong electric fields, which can in turn accelerate particles to high energy, and spark a pair-photon fireball with a near-thermal spectrum (e.g. \citep{Goodman1986}). The observed temperature of such a flare depends sensitively on the baryon load carried in the fireball \citep{Meszaros1993, Nakar2005}. The total fluence of the precursor flare can also provide an estimate for a lower bound on the breaking strain, since both $E_{b}$ and $E_{\rm elastic}$ scale with $\epsilon_{b}^2$.

Motivated by the above calculations, we next explore the range of energetics and timescales expected from a sample of NS crust EOSs, and describe what constraints the observations may provide between them.


{\it Model Exploration.} Using the theoretical framework described above,
we perform the same analysis with the EOSs  APR, SkI6,
SkO, Rs, and Gs, following \citet{Steiner2009} and references therein. 
Results are summarized in Table \ref{tab:1}. 
Our main conclusion is that the energetics
are robust, since they mainly depend on the density of the
crust and displacement eigenfunction, which do not vary greatly between the
models. Much like the toroidal shear modes
\citep{Piro2005a, Steiner2009}, the i-mode frequency is only weakly
dependent on NS mass, but varies with shear speed at the base of the crust. This
implies 
that the resonant excitation of the i-mode would occur at different times
during the merger event.

\begin{table}
\begin{center}
\begin{tabular}{ l c c c c c c } \\ \hline \hline
EOS  & $f_{\rm mode}$ [Hz] &$Q$ & $\Delta E_{\rm max}$ [erg]& $E_b$ [erg] & $\dot{E}_{\rm tidal}$ [erg/s]\\ \hline
SLy4 & 188 & $0.041$ & $5\times10^{50}$ & $5\times 10^{46}$ & $1\times 10^{50}$  \\
 APR & 170 & $0.061$ & $1\times 10^{51}$ & $2\times 10^{46}$ & $9\times 10^{49}$ \\
 SkI6 & 67.3 & 0.017 & $8\times 10^{49}$ & $3\times 10^{45}$ & $1\times10^{48}$  \\
 SkO & 69.1 & 0.053 & $7\times 10^{50}$ & $1\times 10^{46}$ & $1\times 10^{49}$  \\
 Rs & 32.0 & 0.059 & $7\times 10^{50}$ & $1\times 10^{46}$ & $3\times 10^{48}$  \\
 Gs & 28.8 & 0.060 & $8\times 10^{50}$ & $1\times 10^{46}$ & $3\times10^{48}$ \\\hline
\end{tabular}
\caption{Resonant mode properties for the $l=2$ i-mode. The background star is taken to be a $1.4\,M_\odot$ NS, with various equations of state given in \citep{Steiner2009}. The crust/core transition baryon density is fixed to be $n_{\rm t} = 0.065\,{\rm fm}^{-3}$ for each model.}
\label{tab:1}
\end{center}
\end{table}

\begin{figure}[h]
\includegraphics[width=\columnwidth]{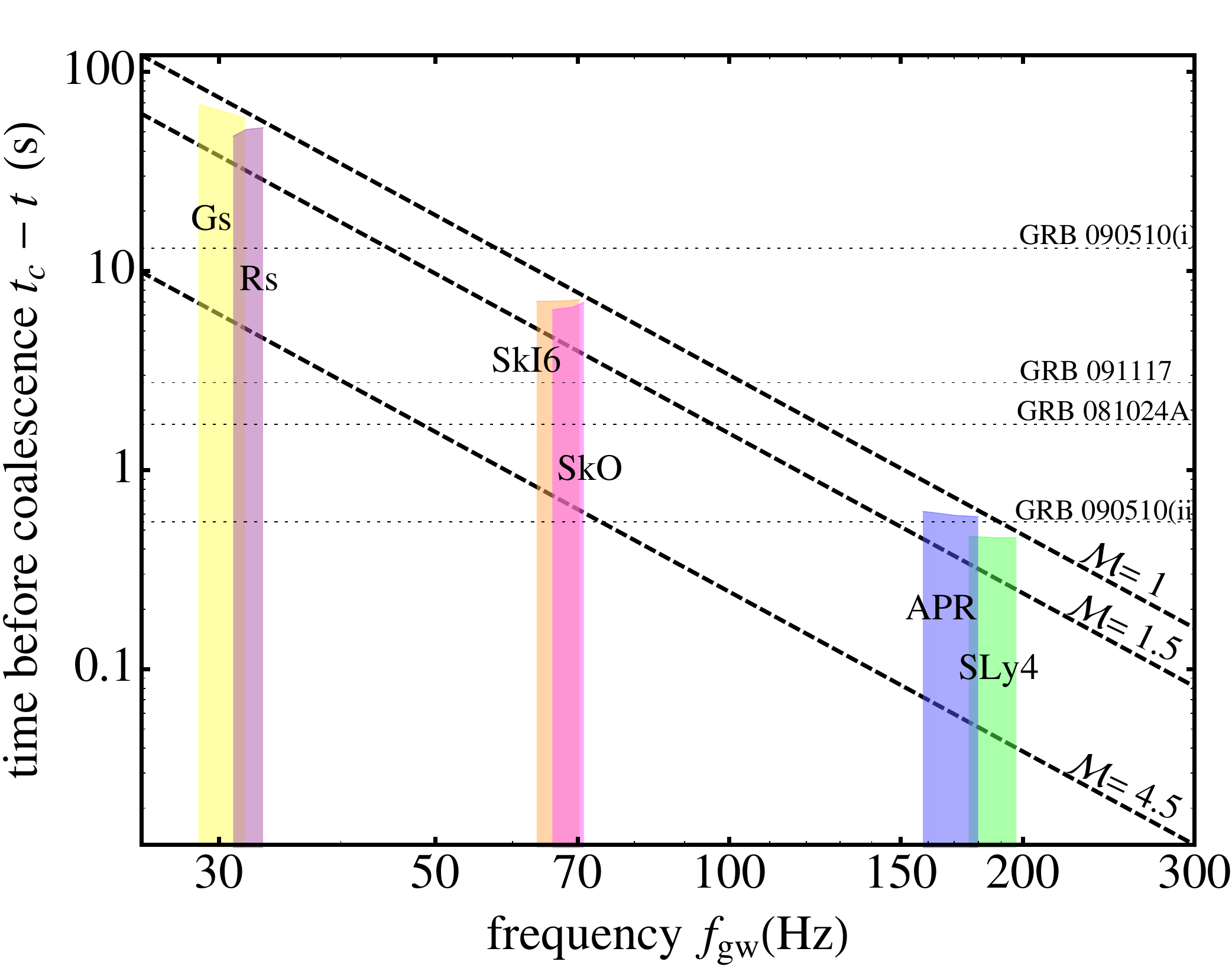}
\caption{\label{fig:constraints} 
The time until PN coalescence ($t_c - t $) as a function of gravitational wave frequency.  The dashed lines show the frequency evolution of inspiraling binaries for different chirp masses $\mathcal{M}$ as labeled in units of $M_\odot$. A given binary moves from left to right in time. The colored columns show the resonance frequencies
$f_{\rm mode}=f_{\rm gw}$ of a set of crust EOSs from \citep{Steiner2009}, over a neutron star mass
range of $1.2\,M_{\odot}$ (higher frequency) to $1.7 M_{\odot}$ (lower frequency). We take $1.2 M_{\odot}$ as the smallest
plausible companion mass, giving an upper bound on the precursor times for
each EOS. NS-NS systems will have chirp masses of $1.0-1.5 M_{\odot}$, and
NS-BH systems with $10-20 M_{\odot}$ BH have chirp masses of $2.7-4.5
M_{\odot}$. The precursor times for the GRBs reported in \citep{Troja2010} are plotted as horizontal dotted lines.}\end{figure}

To quantify the relationship between the crust EOS and the time at which
the i-mode resonance occurs, we plot the i-mode and gravitational wave
frequency versus time until (PN) coalescence, $t_c -t  = 3t_{gw}/8$ (see e.g.
\citep{Blanchet2006}), in Figure \ref{fig:constraints}. The dashed lines
trace the leading order frequency evolution for a given chirp mass $\mathcal{M}$, going
from left to right.  When the dashed line intersects a colored
column, it indicates the time and frequency at which resonance occurs. From
this set of EOSs and $\mathcal{M}$, a wide range of timescale are
possible, from $\lesssim0.1\ {\rm s}$ up to $\approx20\ {\rm s}$ before
merger. Also plotted as horizontal dotted lines are the observed precursor
times reported in \citep{Troja2010}. Although this 
comparison assumes that the main flare is nearly coincident with the
binary coalescence, certain constraints can already be inferred. The
relatively high frequency of the i-mode for the SLy4 EOS means that the
resonance only occurs at late times, close to merger. Only if
$\mathcal{M}\lesssim1M_\odot$ can such a model give timescales similar to
the shortest precursors and the longer precursors may be especially
difficult for this model to replicate. Other EOS models, such as Gs, Rs,
SkI6, and SkO, are largely consistent with the timescale of
precursors, but as a larger sample of precursor observations are made,
diagrams such as this will be useful for constraining EOSs.

A binary with unequal mass NSs may excite two precursor flares separated by
a small time delay, due to the slight difference in the i-mode frequency.
However, the two precursors (13\,s, 0.55\,s) observed in GRB 090510, are
too far separated to both be explained by our resonant shattering model of
precursors, using two NSs with the same EOS.
The 0.55\,s flare may alternatively be evidence of direct crust cracking
\citep{Penner2011} and a delayed main GRB burst, the formation of a
hyper-massive magnetar before collapse into a black hole
\citep{LehnerPrep}, or some other flare mechanism. 



{\it Discussion.} We explored the resonant excitation by tides of a mode that is concentrated at the crust/core boundary of NSs. We demonstrated that the resonance occurs between $\sim0.1-20\ {\rm s}$ prior to merger in NS-NS or NH-NS binaries. Further work remains to be done exploring the details of this model, including the effects of damping on the mode excitation, the effect of more realistic NS structure, and the detailed physics of the magnetospheric emission. However, we have shown that the energetics of the release of mode and elastic energy and the timescale at which the resonance occurs are suggestive of the precursors of sGRBS. Using this theoretical framework we demonstrated that interesting constraints can be placed on the NS crust EOS with comparisons to precursor observations. 

The direct phase change of the gravitational waveform due to the resonant excitation of the mode, $ \Delta \phi\sim (t_{gw}E_{\rm b})/(t_{\rm orbit} E_{\rm orbit}) \sim 10^{-3}\, {\rm rad}$, 
is too small to be directly measured for signal to noise (SNR) $\lo 1000$. However, coincident timing between the $\gamma$-ray burst detectors and the GW detector would allow precise determination of the mode frequency, coalescence time, main burst delay time, and chirp mass. With parameter extraction from the GW inspiral at the detection threshold with SNR $\sim 10$, the dominant error in determining the resonant frequency is due to the uncertainty in the timing of the precursor flare, which is of order the precursor duration. This implies that the mode frequency can be determined to fractional accuracy $\Delta f/f \sim0.1\,{\rm s}/t_{\rm gw} \sim 2\% \, ({\cal M}/1.2)^{5/3} f_{100}^{8/3}$. Such a measurement would allow us to tightly constrain the NS physics and parameters that determine the mode frequency. This is complementary to the constraints given by GW coalescence measurement alone, which are sensitive primarily to the core EOS (e.g. \citep{Hinderer2010,Kyutoku2011}). 

Resonant shattering precursor flares are likely to be fairly isotropic, and thus may be observable even for sGRBs where the main flare is beamed away from the Earth.  Such flares may also be a source of electromagnetic emission for higher mass ratio, lower spin NS-BH mergers where the neutron star does not disrupt to produce a torus and main sGRB flare \citep{Shibata2011, Foucart2011}.


{\it Acknowledgements.} DT, TH, and ALP were supported by the Sherman Fairchild Foundation at Caltech; JSR by 
NSF Grants PHY-0900735 and PHY-1055103; ALP by NASA ATP grant NNX07AH06G and NSF grant AST-0855535; and
RB by NSF Grants PHY 06-53462 \& PHY 09-69857, and NASA Grant NNG05GF71G. 
We thank A. Steiner for providing EOS and compositional tables 
and useful advice. We also thank D. Lai, C. Hirata, B. Metzger, C.
Ott, E. Flanagan, C. Cutler, C. Horowitz, S. Phinney, B. Giacomazzo, P.
Goldreich, A. Lundgren, N. Andersson, C. Gundlach, and the organizers of
the MICRA 2011 workshop for valuable discussion.

\renewcommand{\bibsection}{\section{References}} 

\begin{thebibliography}{35}
\expandafter\ifx\csname natexlab\endcsname\relax\def\natexlab#1{#1}\fi
\expandafter\ifx\csname bibnamefont\endcsname\relax
  \def\bibnamefont#1{#1}\fi
\expandafter\ifx\csname bibfnamefont\endcsname\relax
  \def\bibfnamefont#1{#1}\fi
\expandafter\ifx\csname citenamefont\endcsname\relax
  \def\citenamefont#1{#1}\fi
\expandafter\ifx\csname url\endcsname\relax
  \def\url#1{\texttt{#1}}\fi
\expandafter\ifx\csname urlprefix\endcsname\relax\def\urlprefix{URL }\fi
\providecommand{\bibinfo}[2]{#2}
\providecommand{\eprint}[2][]{\url{#2}}

\bibitem[{\citenamefont{{Nakar}}(2007)}]{nakar:07a}
\bibinfo{author}{\bibfnamefont{E.}~\bibnamefont{{Nakar}}},
  \bibinfo{journal}{Physics Reports} \textbf{\bibinfo{volume}{442}},
  \bibinfo{pages}{166} (\bibinfo{year}{2007}).

\bibitem[{\citenamefont{{Paczynski}}(1986)}]{Paczynski1986}
\bibinfo{author}{\bibfnamefont{B.}~\bibnamefont{{Paczynski}}},
  \bibinfo{journal}{\apjl} \textbf{\bibinfo{volume}{308}}, \bibinfo{pages}{L43}
  (\bibinfo{year}{1986}).

\bibitem[{\citenamefont{{Goodman}}(1986)}]{Goodman1986}
\bibinfo{author}{\bibfnamefont{J.}~\bibnamefont{{Goodman}}},
  \bibinfo{journal}{\apjl} \textbf{\bibinfo{volume}{308}}, \bibinfo{pages}{L47}
  (\bibinfo{year}{1986}).

\bibitem[{\citenamefont{{Abdo} et~al.}(2009)}]{Abdo2009}
\bibinfo{author}{\bibfnamefont{A.~A.} \bibnamefont{{Abdo}}}
  \bibnamefont{et~al.}, \bibinfo{journal}{\nat} \textbf{\bibinfo{volume}{462}},
  \bibinfo{pages}{331} (\bibinfo{year}{2009}).

\bibitem[{\citenamefont{Troja et~al.}(2010)\citenamefont{Troja, Rosswog, and
  Gehrels}}]{Troja2010}
\bibinfo{author}{\bibfnamefont{E.}~\bibnamefont{Troja}},
  \bibinfo{author}{\bibfnamefont{S.}~\bibnamefont{Rosswog}}, \bibnamefont{and}
  \bibinfo{author}{\bibfnamefont{N.}~\bibnamefont{Gehrels}},
  \bibinfo{journal}{\apj} \textbf{\bibinfo{volume}{723}}, \bibinfo{pages}{1711}
  (\bibinfo{year}{2010}).

\bibitem[{\citenamefont{{Burlon} et~al.}(2009)}]{Burlon2009}
\bibinfo{author}{\bibfnamefont{D.}~\bibnamefont{{Burlon}}}
  \bibnamefont{et~al.}, \bibinfo{journal}{\aap} \textbf{\bibinfo{volume}{505}},
  \bibinfo{pages}{569} (\bibinfo{year}{2009}).

\bibitem[{\citenamefont{{Koshut} et~al.}(1995)}]{Koshut1995}
\bibinfo{author}{\bibfnamefont{T.~M.} \bibnamefont{{Koshut}}}
  \bibnamefont{et~al.}, \bibinfo{journal}{\apj} \textbf{\bibinfo{volume}{452}},
  \bibinfo{pages}{145} (\bibinfo{year}{1995}).

\bibitem[{\citenamefont{{Hansen} and {Lyutikov}}(2001)}]{Hansen2001}
\bibinfo{author}{\bibfnamefont{B.~M.~S.} \bibnamefont{{Hansen}}}
  \bibnamefont{and}
  \bibinfo{author}{\bibfnamefont{M.}~\bibnamefont{{Lyutikov}}},
  \bibinfo{journal}{\mnras} \textbf{\bibinfo{volume}{322}},
  \bibinfo{pages}{695} (\bibinfo{year}{2001}).

\bibitem[{\citenamefont{Kochanek}(1992)}]{Kochanek1992b}
\bibinfo{author}{\bibfnamefont{C.~S.} \bibnamefont{Kochanek}},
  \bibinfo{journal}{\apj} \textbf{\bibinfo{volume}{398}}, \bibinfo{pages}{234}
  (\bibinfo{year}{1992}).

\bibitem[{\citenamefont{{Penner} et~al.}(2011)\citenamefont{{Penner},
  Andersson, Jones, Samuelsson, and Hawke}}]{Penner2011}
\bibinfo{author}{\bibfnamefont{J.}~\bibnamefont{{Penner}}},
  \bibinfo{author}{\bibfnamefont{N.}~\bibnamefont{Andersson}},
  \bibinfo{author}{\bibfnamefont{D.}~\bibnamefont{Jones}},
  \bibinfo{author}{\bibfnamefont{L.}~\bibnamefont{Samuelsson}},
  \bibnamefont{and} \bibinfo{author}{\bibfnamefont{I.}~\bibnamefont{Hawke}},
  \bibinfo{journal}{arXiv:1109.5041, Astrophs. J. Lett. submitted}
  (\bibinfo{year}{2011}).

\bibitem[{\citenamefont{Horowitz and Kadau}(2009)}]{Horowitz2009b}
\bibinfo{author}{\bibfnamefont{C.}~\bibnamefont{Horowitz}} \bibnamefont{and}
  \bibinfo{author}{\bibfnamefont{K.}~\bibnamefont{Kadau}},
  \bibinfo{journal}{\prl} \textbf{\bibinfo{volume}{102}}
  (\bibinfo{year}{2009}).

\bibitem[{\citenamefont{Damour and Nagar}(2009)}]{Damour2009a}
\bibinfo{author}{\bibfnamefont{T.}~\bibnamefont{Damour}} \bibnamefont{and}
  \bibinfo{author}{\bibfnamefont{A.}~\bibnamefont{Nagar}},
  \bibinfo{journal}{\prd} \textbf{\bibinfo{volume}{80}} (\bibinfo{year}{2009}).

\bibitem[{\citenamefont{Vines et~al.}(2011)\citenamefont{Vines, Hinderer, and
  Flanagan}}]{Vines2011}
\bibinfo{author}{\bibfnamefont{J.}~\bibnamefont{Vines}},
  \bibinfo{author}{\bibfnamefont{T.}~\bibnamefont{Hinderer}}, \bibnamefont{and}
  \bibinfo{author}{\bibfnamefont{E.~E.} \bibnamefont{Flanagan}},
  \bibinfo{journal}{\prd} \textbf{\bibinfo{volume}{83}},
  \bibinfo{pages}{084051} (\bibinfo{year}{2011}).

\bibitem[{\citenamefont{{Misner} et~al.}(1973)\citenamefont{{Misner}, {Thorne},
  and {Wheeler}}}]{MTW}
\bibinfo{author}{\bibfnamefont{C.~W.} \bibnamefont{{Misner}}},
  \bibinfo{author}{\bibfnamefont{K.~S.} \bibnamefont{{Thorne}}},
  \bibnamefont{and} \bibinfo{author}{\bibfnamefont{J.~A.}
  \bibnamefont{{Wheeler}}}, \emph{\bibinfo{title}{{Gravitation}}}
  (\bibinfo{publisher}{San Francisco, W.H.~Freeman and Co.},
  \bibinfo{year}{1973}).

\bibitem[{\citenamefont{Steiner and Watts}(2009)}]{Steiner2009}
\bibinfo{author}{\bibfnamefont{A.}~\bibnamefont{Steiner}} \bibnamefont{and}
  \bibinfo{author}{\bibfnamefont{A.}~\bibnamefont{Watts}},
  \bibinfo{journal}{\prl} \textbf{\bibinfo{volume}{103}}
  (\bibinfo{year}{2009}).

\bibitem[{\citenamefont{{Lai}}(1994)}]{Lai1994}
\bibinfo{author}{\bibfnamefont{D.}~\bibnamefont{{Lai}}},
  \bibinfo{journal}{\mnras} \textbf{\bibinfo{volume}{270}},
  \bibinfo{pages}{611} (\bibinfo{year}{1994}).

\bibitem[{\citenamefont{Shibata}(1994)}]{Shibata1994}
\bibinfo{author}{\bibfnamefont{M.}~\bibnamefont{Shibata}},
  \bibinfo{journal}{Prog. Theor. Phys.} \textbf{\bibinfo{volume}{91}},
  \bibinfo{pages}{871} (\bibinfo{year}{1994}).

\bibitem[{\citenamefont{{Reisenegger} and {Goldreich}}(1994)}]{Reisenegger1994}
\bibinfo{author}{\bibfnamefont{A.}~\bibnamefont{{Reisenegger}}}
  \bibnamefont{and}
  \bibinfo{author}{\bibfnamefont{P.}~\bibnamefont{{Goldreich}}},
  \bibinfo{journal}{\apj} \textbf{\bibinfo{volume}{426}}, \bibinfo{pages}{688}
  (\bibinfo{year}{1994}).

\bibitem[{\citenamefont{McDermott et~al.}(1988)\citenamefont{McDermott, van
  Horn, and Hansen}}]{McDermott1988}
\bibinfo{author}{\bibfnamefont{P.~N.} \bibnamefont{McDermott}},
  \bibinfo{author}{\bibfnamefont{H.~M.} \bibnamefont{van Horn}},
  \bibnamefont{and} \bibinfo{author}{\bibfnamefont{C.~J.}
  \bibnamefont{Hansen}}, \bibinfo{journal}{\apj}
  \textbf{\bibinfo{volume}{325}}, \bibinfo{pages}{725} (\bibinfo{year}{1988}).

\bibitem[{\citenamefont{Strohmayer et~al.}(1991)\citenamefont{Strohmayer, van
  Horn, Ogata, Iyetomi, and Ichimaru}}]{Strohmayer1991}
\bibinfo{author}{\bibfnamefont{T.}~\bibnamefont{Strohmayer}},
  \bibinfo{author}{\bibfnamefont{H.~M.} \bibnamefont{van Horn}},
  \bibinfo{author}{\bibfnamefont{S.}~\bibnamefont{Ogata}},
  \bibinfo{author}{\bibfnamefont{H.}~\bibnamefont{Iyetomi}}, \bibnamefont{and}
  \bibinfo{author}{\bibfnamefont{S.}~\bibnamefont{Ichimaru}},
  \bibinfo{journal}{\apj} \textbf{\bibinfo{volume}{375}}, \bibinfo{pages}{679}
  (\bibinfo{year}{1991}).

\bibitem[{\citenamefont{{Harding} and {Lai}}(2006)}]{Harding2006}
\bibinfo{author}{\bibfnamefont{A.~K.} \bibnamefont{{Harding}}}
  \bibnamefont{and} \bibinfo{author}{\bibfnamefont{D.}~\bibnamefont{{Lai}}},
  \bibinfo{journal}{Reports on Progress in Physics}
  \textbf{\bibinfo{volume}{69}}, \bibinfo{pages}{2631} (\bibinfo{year}{2006}).

\bibitem[{\citenamefont{Landau and Lifshitz}(1959)}]{Landau}
\bibinfo{author}{\bibfnamefont{L.~D.} \bibnamefont{Landau}} \bibnamefont{and}
  \bibinfo{author}{\bibfnamefont{E.~M.} \bibnamefont{Lifshitz}},
  \emph{\bibinfo{title}{Theory of elasticity}} (\bibinfo{publisher}{London,
  Pergamon}, \bibinfo{year}{1959}).

\bibitem[{\citenamefont{Blaes et~al.}(1989)\citenamefont{Blaes, Blandford,
  Goldreich, and Madau}}]{Blaes1989}
\bibinfo{author}{\bibfnamefont{O.}~\bibnamefont{Blaes}},
  \bibinfo{author}{\bibfnamefont{R.}~\bibnamefont{Blandford}},
  \bibinfo{author}{\bibfnamefont{P.}~\bibnamefont{Goldreich}},
  \bibnamefont{and} \bibinfo{author}{\bibfnamefont{P.}~\bibnamefont{Madau}},
  \bibinfo{journal}{\apj} \textbf{\bibinfo{volume}{343}}, \bibinfo{pages}{839}
  (\bibinfo{year}{1989}).

\bibitem[{\citenamefont{{Bildsten} and {Cutler}}(1992)}]{BildstenCutler1992}
\bibinfo{author}{\bibfnamefont{L.}~\bibnamefont{{Bildsten}}} \bibnamefont{and}
  \bibinfo{author}{\bibfnamefont{C.}~\bibnamefont{{Cutler}}},
  \bibinfo{journal}{\apj} \textbf{\bibinfo{volume}{400}}, \bibinfo{pages}{175}
  (\bibinfo{year}{1992}).

\bibitem[{\citenamefont{{Thompson} and {Duncan}}(1995)}]{Thompson1995}
\bibinfo{author}{\bibfnamefont{C.}~\bibnamefont{{Thompson}}} \bibnamefont{and}
  \bibinfo{author}{\bibfnamefont{R.}~\bibnamefont{{Duncan}}},
  \bibinfo{journal}{\mnras} \textbf{\bibinfo{volume}{275}},
  \bibinfo{pages}{255} (\bibinfo{year}{1995}).

\bibitem[{\citenamefont{{Thompson} and {Blaes}}(1998)}]{Thompson1998}
\bibinfo{author}{\bibfnamefont{C.}~\bibnamefont{{Thompson}}} \bibnamefont{and}
  \bibinfo{author}{\bibfnamefont{O.}~\bibnamefont{{Blaes}}},
  \bibinfo{journal}{\prd} \textbf{\bibinfo{volume}{57}}, \bibinfo{pages}{3219}
  (\bibinfo{year}{1998}).

\bibitem[{\citenamefont{{Meszaros} et~al.}(1993)\citenamefont{{Meszaros},
  {Laguna}, and {Rees}}}]{Meszaros1993}
\bibinfo{author}{\bibfnamefont{P.}~\bibnamefont{{Meszaros}}},
  \bibinfo{author}{\bibfnamefont{P.}~\bibnamefont{{Laguna}}}, \bibnamefont{and}
  \bibinfo{author}{\bibfnamefont{M.~J.} \bibnamefont{{Rees}}},
  \bibinfo{journal}{\apj} \textbf{\bibinfo{volume}{415}}, \bibinfo{pages}{181}
  (\bibinfo{year}{1993}).

\bibitem[{\citenamefont{{Nakar} et~al.}(2005)\citenamefont{{Nakar}, {Piran},
  and {Sari}}}]{Nakar2005}
\bibinfo{author}{\bibfnamefont{E.}~\bibnamefont{{Nakar}}},
  \bibinfo{author}{\bibfnamefont{T.}~\bibnamefont{{Piran}}}, \bibnamefont{and}
  \bibinfo{author}{\bibfnamefont{R.}~\bibnamefont{{Sari}}},
  \bibinfo{journal}{\apj} \textbf{\bibinfo{volume}{635}}, \bibinfo{pages}{516}
  (\bibinfo{year}{2005}).

\bibitem[{\citenamefont{Piro}(2005)}]{Piro2005a}
\bibinfo{author}{\bibfnamefont{A.~L.} \bibnamefont{Piro}},
  \bibinfo{journal}{\apj} \textbf{\bibinfo{volume}{634}}, \bibinfo{pages}{L153}
  (\bibinfo{year}{2005}).

\bibitem[{\citenamefont{Blanchet}(2006)}]{Blanchet2006}
\bibinfo{author}{\bibfnamefont{L.}~\bibnamefont{Blanchet}},
  \bibinfo{journal}{Living Reviews in Relativity} \textbf{\bibinfo{volume}{9}}
  (\bibinfo{year}{2006}).

\bibitem[{\citenamefont{{Lehner}}(2011)}]{LehnerPrep}
\bibinfo{author}{\bibfnamefont{L.}~\bibnamefont{{Lehner}}},
  \bibinfo{journal}{private communication}  (\bibinfo{year}{2011}).

\bibitem[{\citenamefont{Hinderer et~al.}(2010)\citenamefont{Hinderer, Lackey,
  Lang, and Read}}]{Hinderer2010}
\bibinfo{author}{\bibfnamefont{T.}~\bibnamefont{Hinderer}},
  \bibinfo{author}{\bibfnamefont{B.}~\bibnamefont{Lackey}},
  \bibinfo{author}{\bibfnamefont{R.}~\bibnamefont{Lang}}, \bibnamefont{and}
  \bibinfo{author}{\bibfnamefont{J.~S.} \bibnamefont{Read}},
  \bibinfo{journal}{\prd} \textbf{\bibinfo{volume}{81}}, \bibinfo{pages}{1}
  (\bibinfo{year}{2010}).

\bibitem[{\citenamefont{Kyutoku et~al.}(2011)\citenamefont{Kyutoku, Okawa,
  Shibata, and Taniguchi}}]{Kyutoku2011}
\bibinfo{author}{\bibfnamefont{K.}~\bibnamefont{Kyutoku}},
  \bibinfo{author}{\bibfnamefont{H.}~\bibnamefont{Okawa}},
  \bibinfo{author}{\bibfnamefont{M.}~\bibnamefont{Shibata}}, \bibnamefont{and}
  \bibinfo{author}{\bibfnamefont{K.}~\bibnamefont{Taniguchi}},
  \bibinfo{journal}{Phys. Rev. D} \textbf{\bibinfo{volume}{84}},
  \bibinfo{pages}{064018} (\bibinfo{year}{2011}).

\bibitem[{\citenamefont{{Shibata} and {Taniguchi}}(2011)}]{Shibata2011}
\bibinfo{author}{\bibfnamefont{M.}~\bibnamefont{{Shibata}}} \bibnamefont{and}
  \bibinfo{author}{\bibfnamefont{K.}~\bibnamefont{{Taniguchi}}},
  \bibinfo{journal}{Living Reviews in Relativity}
  \textbf{\bibinfo{volume}{14}}, \bibinfo{pages}{6} (\bibinfo{year}{2011}).

\bibitem[{\citenamefont{{Foucart} et~al.}(2011)\citenamefont{{Foucart}, {Duez},
  {Kidder}, {Scheel}, {Szilagyi}, and {Teukolsky}}}]{Foucart2011}
\bibinfo{author}{\bibfnamefont{F.}~\bibnamefont{{Foucart}}},
  \bibinfo{author}{\bibfnamefont{M.~D.} \bibnamefont{{Duez}}},
  \bibinfo{author}{\bibfnamefont{L.~E.} \bibnamefont{{Kidder}}},
  \bibinfo{author}{\bibfnamefont{M.~A.} \bibnamefont{{Scheel}}},
  \bibinfo{author}{\bibfnamefont{B.}~\bibnamefont{{Szilagyi}}},
  \bibnamefont{and} \bibinfo{author}{\bibfnamefont{S.~A.}
  \bibnamefont{{Teukolsky}}}, \bibinfo{journal}{arXiv:1111.1677, \prd ~submitted}
  (\bibinfo{year}{2011}).

\end{thebibliography}

\bibliographystyle{apsrev}

\end{document}